\documentclass{iopart}[12]
\usepackage{amsfonts}
\usepackage{graphicx}
\usepackage{subfigure}
\pdfoutput=1 

\begin{document}

\renewcommand{\baselinestretch}{0.833}
\newcommand{\beq}{\begin{equation}}
\newcommand{\eeq}{\end{equation}}
\newcommand{\beqar}{\begin{eqnarray}}
\newcommand{\eeqar}{\end{eqnarray}}

\newcommand{\gsim}{\stackrel{>}{\sim}}
\newcommand{\spc}{\mbox{ }}
\newcommand{\dspc}{\mbox{  }}

\title[Robust Gravitational Wave Burst Detection...]
{Robust Gravitational Wave Burst Detection\\ and Source Localization in a Network of Interferometers
Using Cross Wigner Spectra}

\author{Rocco P Croce$^1$, Vincenzo Pierro$^1$, Fabio Postiglione$^2$, Maria Principe$^1$ and Innocenzo M Pinto$^1$}
\address{$^1$WavesGroup, University of Sannio at Benevento, Italy, INFN and LSC,
$^2$ Dept. of Electronic and Computer Engineering, University of Salerno, Italy}
\eads{\mailto{principe@unisannio.it}}
\date{LIGO-P1100042-v1}
\pacno{04.80.Nn, 07.05.Kf, 95.55.Sz}

\begin{abstract}
We discuss  a fast cross-Wigner transform based technique for detecting gravitational wave bursts, 
and estimating the direction of arrival, using a network of (three) non co-located interferometric detectors. 
The performances of the detector as a function of signal strength and source location, 
and the accuracy of the direction of arrival estimation are investigated by numerical simulations.
The robustness of the method against instrumental glitches is illustrated. 
\end{abstract}


\section{Introduction}
\label{sect:intro}

The next generation of  interferometric detectors, 
of gravitational waves (henceforth GW)  
including AdLIGO \cite{AdLIGO}, AdVirgo \cite{AdVirgo} and GEO-HF \cite{superGEO},
hopefully to be followed soon by LCGT \cite{LCGT} and ACIGA \cite{ACIGA},
and eventually by ET \cite{ET}, 
is expected to observe tens of events per year, 
opening the way to gravitational wave astronomy \cite{GWBs}.
Identifying the direction of arrival (henceforth DOA) of the signals,
and retrieving their shapes, 
will be a primary task in reconstructing the physics of the sources
and their environments.\\
The possibility of retrieving the DOA from {\it independent} estimates of the signal arrival time 
at {\it each} detector was first suggested in \cite{Boulanger}, and further discussed in Saulson seminal book \cite{Saulson_book}.  
It was shown that three-interferometers are sufficient to retrieve the DOA up to a mirror-image  ambiguity 
which can be solved in principle from knowledge of the detectors' directional responses. 
This method, often referred to as {\it triangulation} was further elaborated by Sylvestre \cite{Sylvestre_03}, 
Cavalier et al. \cite{Cavalier_06}, and Merkovitz et al. \cite{Zanolin_08}.
In \cite{Cavalier_06} a Gaussian distribution was assumed for the (independent) arrival time estimation errors,
and a $\chi^2$ minimization algorithm was accordingly proposed for retrieving the DOA,
in the maximum likelihood spirit. 
In \cite{Zanolin_08} it was shown that this method is affected by 
a systematic bias in the estimated DOA, a possible technique for removing the bias was discussed,
and amplitude consistency tests for removing the mirror-image ambiguity were suggested.
Fairhurst developed a similar analysis of the effect of arrival time estimation errors 
on the DOA estimation accuracy, 
for the special case of chirping signals, including waveform and calibration errors \cite{Fairh_09}, \cite{Fairh_11}.\\
DOA estimation algorithms are already implemented in the {\it coherent} 
LIGO-Virgo pipelines for GW burst (henceforth GWB) detection  
\cite{Xpipe}, \cite{Klimenko_11}.
DOA estimation in {\it coherent} network data analysis,
was studied  first by Krolak and Jaranowski \cite{Kro_Jar_94}, and then by Pai et al. \cite{Pai_Dhur}, 
as part of  the waveform parameter estimation problem,  with specific reference 
to chirping waveforms from coalescing binaries, in a Gaussian noise background.
The conceptual foundations of coherent data analysis for unmodeled waveforms were laid out
by Flanagan and Hughes \cite{Flan_Hug}, and further developed by Klimenko et al. \cite{Klim1}-\cite{Rakh}.  
G\"{u}rsel and Tinto \cite{GurTin} first suggested the possibility of retrieving the DOA for un modeled signals
using null-streams. This concept was  analyzed in depth by Schutz and Wen \cite{Wen_Schutz_05}, and
further exploited by Chatterji et al. \cite{Shurov}.
A Fisher-matrix based analysis of arrival time estimation error in coherent network detection 
of modeled as well as {\it unmodeled} signals was made in Wen et al. \cite{Wen_Fan_Chen}.\\
In this paper we capitalize on the time-shift and localization properties of the cross-Wigner-Ville
(henceforth XWV)
transform to introduce a new and conceptually simple 
GWB detection and DOA reconstruction algorithm,
using a network of non co-located interferometric detectors.\\
The Wigner-Ville transform is a well known powerful tool for the analysis of non-stationary
signals \cite{XWV1}, whose potential in GW data analysis, has been highlighted by several
Authors, under different perspectives \cite{Feo_etal}-\cite{Chass}.
Here we suggest its possible use as an effective tool 
for detecting GWBs, and estimating their DOA, 
which offers nice features in terms of performance, 
robustness against spurious instrumental/environmental transients (glitches).\\
Instead of using {\it independent} estimates of the arrival times at {\it each} detector, 
our DOA estimator uses data from (all) detector {\it pairs} to estimate the needed propagation delays. 
In addition, it also provides an effective detection statistic, combining the data from {\it all} detectors in the network, 
at a remarkably light computational cost.\\
DOA reconstruction from arrival-time delay estimation in a network of sensors 
is a well known problem in the technical Literature on Acoustics and Radar  (see, e.g.,  \cite{Berdugo} for a broad review).
The standard method for time-delay estimation in Gaussian noise 
is (generalized) cross-correlation \cite{Knapp}, which is known 
to  perform reasonably well for relatively large signal to noise ratios \cite{Fert_Sjo}.
Remarkably, the correlation-based estimator 
offers worse performances compared to the XWV in the present context, 
as shown in Sect. 5.2\\
This paper is  organized as follows. In Sect.  2 we introduce the XWV transform, and recall 
its time-shift properties, which are illustrated  for the simplest case of 
sine-Gaussian (henceforth SG) GWBs. In the same section 
we recall the relationship between arrival time delays and DOA.
In Sect. 3, we  illustrate the proposed XWV transform based DOA reconstruction algorithm. 
In Sect. 4 we discuss the effect of noise in the data, 
and the related DOA reconstruction uncertainties.
In Sect. 5 we present the results of extensive numerical simulations, 
aimed at characterizing the performance of our XWV based algorithm both as a detector and
as a DOA estimator. The simulations are based on SG-GWBs, but
the case of more realistic waveforms (including Dimmelmeier and binary merger waveforms)
is also discussed.
In Sect. 6  we include a short discussion  of the robustness of the proposed algorithm against 
instrumental/environmental transients (glitches).
Conclusions follow under Sect. 7.

\section{Rationale. From Cross-Wigner-Ville Transforms to DOAs}
\label{sect:XWIG}
%
In this section we recall a relevant property of the XWV transform,
and illustrate it using ideal (sine-Gaussian)  waveforms.
We further recall the relationship between the arrival time delays and the
DOA, for a 3-detectors network, with special reference to the LIGO-Virgo
Observatory.

\subsection{Cross-Wigner-Ville Transforms}
%
The XWV transform built from
two (analytic, complex) signals 
$\tilde{x}_{1,2}$ is given by \cite{XWV2}, \cite{XWV3}: 
\beq
W_{12}(t,f)=\int_{-\infty}^{\infty} d\theta
\mbox{ }
\tilde{x}_1^*
\left(
t-\frac{\theta}{2}
\right)
\tilde{x}_2
\left(
t+\frac{\theta}{2}
\right)
\exp
\left(
-2\pi\imath f \theta
\right)
\label{eq:crosswig}
\eeq
where $*$ denotes complex conjugation.
We recall that the so called {\it analytic signal} 
corresponding to a generic  real-valued waveform $x(t)$ is 
\beq
\tilde{x}(t)=x(t)+\imath{\cal H}[x](t),
\label{eq:an_sig}
\eeq
where ${\cal H}[x](t)$ is the Hilbert transform \cite{ansig}. For $x_1(t)=x_2(t)=x(t)$,
eq. (\ref{eq:crosswig}) reduces to the well known Wigner-Ville transform of $x(t)$. \\
%
\subsection{Time-Shift Property of Cross-Wigner-Ville Transform}
%
Let $T_{\theta}$ the time-shift operator, such that
\beq
T_{\theta}[x]=x(t-\theta).
\label{eq:shift}
\eeq
The following property of the XWV transform is easily proved:
\beq
W_{T_{\theta_1}[x_1],T_{\theta_2}[x_2]}(t,f)=\exp[-2\pi\imath f(\theta_2-\theta_1)]
W_{x_1,x_2}\left[t-\frac{\theta_2+\theta_1}{2},f\right].
\label{eq:shift_Th}
\eeq
Hence, if $x_1$ and $x_2$ are {\it the same} waveform $x(t)$, except for having 
different amplitudes, and different time-shift (delays), one accordingly has:
\beq
\left|
W_{T_{\theta_1}[x_1],T_{\theta_2}[x_2]}(t,f)
\right|=
C\left|
W_{x,x}\left[t-\frac{\theta_2+\theta_1}{2},f\right]
\right|.
\label{eq:WVpeak}
\eeq
where $C$ is an irrelevant (positive) constant.
%
\subsubsection{Sine Gaussian GWBs}
%
To illustrate the practical significance of eq. (\ref{eq:WVpeak}) in the context of GW detection of
unmodeled transients using a network of interferometers, we shall refer here
to  SG waveforms,
which have been widespreadly used to model GWBs. More realistic
transient waveforms will be considered in Section 5.3.
Consider two SG waveforms, with common carrier frequency $f_0$,
time spread $T$,  and initial phases $\phi_0$, 
peaked at $t_{1,2}$, with amplitudes $A_{1,2}$, respectively, viz.: 
\beq
x_i(t)=A_i\cos\left[2\pi f_0(t-t_i)+\phi_0\right]
\exp[-(t-t_i)^2/T^2],\mbox{  }i=1,2
\label{eq:SGs}
\eeq
Under the assumption $f_0T \gg 1$, the analytic counterparts of (\ref{eq:SGs}) 
are asymptotically given by
\beq
\tilde{x}_i(t) \sim A_i\exp\left[2\pi\imath  f_0(t-t_i)+\imath\phi_0\right]
\exp[-(t-t_i)^2/T^2],\mbox{  }i=1,2
\label{eq:ASGs}
\eeq
The XWV spectrum, eq. (\ref{eq:crosswig}) between $\tilde{x}_1$ and $\tilde{x}_2$ 
can be computed in closed form, yielding
$$
W_{12}(t,f)=W_{21}(t,f)=(2\pi)^{1/2} A_1 A_2 T
\exp\left[-2\pi^2 T^2 (f-f_0)^2\right]\cdot
$$
\beq
\cdot\exp\left[-\frac{2}{T^2} \left(t-\frac{t_1+t_2}{2}\right)^2\right]
\exp\left[-2\imath\pi f(t_1-t_2) \right].
\eeq 
It is seen that $|W_{12}(t,f)|$  is {\it peaked} at 
\beq
t=\frac{t_1+t_2}{2},
\mbox{   }
f=f_0.
\label{eq:peakpos}
\eeq
\subsubsection{Realistic Waveforms}
\label{sect:SNB}
%
The above {\it peak localization} property of the XWV holds true not only for SG waveforms, 
but essentially for {\it all} waveforms modeled by oscillatory transients with {\it unimodal} envelope,
provided the product between the (instantaneous) carrier frequency and the envelope duration
is a large number.
Under this respect, the SG waveform is a kind of (worst) limiting case in view of its {\it minimal spread}
property in the time frequency plane. Indeed, the localization property can be {\it more} marked for other
transient waveforms, like, e.g., those numerically generated for supernovas or mergers, 
as  discussed in Section 5.3.
%
\subsection{XWV Spectra, Delays and DOAs}
\label{sect:XWV_to_DOA}
%
Let us confine for simplicity to the relevant case of the  LIGO-Virgo network, which
consists of the three large-baseline detectors located at Livingston LA (USA), Hanford WA (USA) and
Cascina (Italy), henceforth denoted as  L1, H1, and V, and labeled by the suffix $i=1,2,3$,
respectively.
In the presence of a GWB, in view of eq. (\ref{eq:peakpos}), the three XWV spectra computed 
from the data gathered by the LIGO-Virgo network interferometers will be (scaled) replicas of the
Wigner-Ville transform of the observed GWB, exhibiting magnitude peaks at\footnote{
We implicitly assume the interferometers' transfer functions as being frequency independent
throughout the useful band of the sought signals.}
%
\beq
t=T_{ij}=\frac{\tau_i+\tau_j}{2},
\mbox{ }
f=f_{ij}=f_0,
\mbox{ }
\{i,j\}=\{1,2\},\{1,3\},\{2,3\},
\label{eq:LVpeaks}
\eeq
where $\tau_i$ is the GWB arrival time at detector-$i$.
Knowledge of the three $T_{ij}$ from the corresponding XWV peaks allows to retrieve in principle
two independent arrival-time delays, e.g.,
\beq
t_{13}=\tau_1-\tau_3=2(T_{12}-T_{23}), 
\mbox{ }
t_{23}=\tau_2-\tau_3=2(T_{12}-T_{13}), 
\label{eq:peaks}
\eeq
from which the DOA, and hence the source location 
on the celestial sphere can be uniquely inferred, as shown in the next subsection.
%
\subsubsection{DOA from Delays}
\label{sect:DOA_from_del}
%
The DOA is easily retrieved from the arrival-time delays
using the reference system sketched in Figure 1, whose origin
is the circumcenter $O$ of the triangle whose vertexes are: 
(1) LIGO-Livingston (L1),  
(2) LIGO Hanford (H1), and
(3) and Virgo (V), 
and whose $x$-axis goes, e.g., through L1.
In this reference system the three detectors have spherical polar coordinates 
\beq
(\vartheta_i=\pi/2,\varphi=\varphi_i),
\mbox{ }
i=1,2,3
\label{eq:det_ang}
\eeq
and are located at:
\beq
\vec{r}_i=R(\hat{u}_x\cos\varphi_i +\hat{u}_y\sin\varphi_i)
\mbox{ },
i=1,2,3
\label{eq:ri}
\eeq
where $\varphi_1=0$ by construction, and $R$ is the radius of the circumference 
through L1, H1 and V. 
Let the source polar coordinates and vector position be 
$\vartheta=\vartheta_s,\varphi=\varphi_s$ and
\beq
\vec{r}=\rho(\sin\vartheta\cos\varphi \hat{u}_x+\sin\vartheta\sin\varphi \hat{u}_y+\cos\vartheta\hat{u}_z),
\label{eq:source}
\eeq
respectively, where $\rho$ is the distance of the source from $O$.
Under the obvious assumption where $\rho \gg R$, one has
\beq
\left|
\vec{r}-\vec{r}_i
\right|
\sim
\rho-R\sin\vartheta_s\cos(\varphi_s-\varphi_i)
\mbox{ },
i=1,2,3.
\label{eq:rij}
\eeq
whence the delays between the wavefront arrival times at the detectors are
\beq
\left.
\begin{array}{l}
t_{ij}=\tau_i-\tau_j=c^{-1}R\sin\vartheta_s
\left[
\cos(\varphi_s-\varphi_j)
-\cos(\varphi_s-\varphi_i)
\right],\\
\hspace*{27pt}\{i,j\}=\{1,2\},\{1,3\},\{2,3\},
\end{array}
\right.
\label{eq:delays}
\eeq
where $c$ is the speed of  light in vacuum.
From the ratio $\xi=t_{13}/t_{23}$ one may accordingly retrieve $\varphi_s$ as follows,
\beq
\varphi_s=-\tan^{-1}
\left[
\frac{(1-\xi)\cos\varphi_3+\xi\cos\varphi_2-\cos\varphi_1}
{(1-\xi)\sin\varphi_3+\xi\sin\varphi_2-\sin\varphi_1}
\right].
\eeq
Once $\varphi_s$ has been computed, it can be used in (any of) eqs.
(\ref{eq:delays}) to retrieve $\vartheta_s$.
Note that the delays (\ref{eq:delays}) do {\it not} change upon letting 
$\vartheta_s\longrightarrow\pi-\vartheta_s$,
yielding the source mirror image w.r.t. to the detectors' plane.
The above mirror-image ambiguity in a 3-detectors network is well known\footnote{
The mirror-image ambiguity can be resolved, in principle, 
from knowledge of the detectors'  pattern functions, 
featuring {\it different} responses in the
$\vartheta=\vartheta_s$ and $\vartheta=\pi-\vartheta_s$ directions.}
\cite{Saulson_book}.
%
\subsection{XWV Spectra of Noise}
%
As a preparation for the next sections, it is  important 
to characterize the key features of the XWV spectrum of independent, pure stationary Gaussian noise streams
(the effect of instrumental transients, aka glitches, will be discussed in Sect. \ref{sec:Glitches}).
In this case, the time-frequency levels in the XWV spectrum will be random, 
and their statistical distribution, in view of the assumed noise stationarity, will be the same for all (discrete) times.\\
The first two moments of the above distribution can be computed analytically with relative ease \cite{Stankovic}.
In particular, for all (discrete) frequencies the average value is zero, 
and the variance exhibits a piecewise linear dependence on frequency,
as sketched in Figure 2. The  maximum variance occurs at  $f=f_s/2$,  where $f_s$ is the sampling frequency,
and its value depends on the details of the XWV implementation (size, windowing), 
and the noise level in the data streams (see Appendix for details).
It is thus expedient to {\it equalize} the XWV time-frequency levels, 
so as to obtain a uniform (flat) XWV spectrum for  pure-noise data streams.
To this end, we merely scale the XWV level in each time frequency pixel  
to the (computed) standard deviation of the XWV level in that pixel. \\ 
%
\section{Estimating DOAs from Discrete XWV Spectra of Noisy Data}
\label{sect:DOA}
%
In practice,  the XWV spectra will be computed in {\it discrete} form \cite{XWV2}, 
yielding  two-dimensional (complex) arrays, rather than continuous time-frequency functions over $\mathbb{R}^2$.\\ 
To minimize the effect of time-discretization error
it is convenient to estimate the independent delays corresponding 
to the {\it largest} available baselines, i.e., in our case, $t_{13}$ (L1-V)
and $t_{23}$ (H1-V).\\
Also, in the presence of noise eqs.  (\ref{eq:peaks})  used in (\ref{eq:delays}) will provide a mere {\it estimate} of the DOA,
whose goodness will basically depend on the available signal to noise ratio, 
which affects the accuracy whereby the XWV peaks can be identified.\\
A simple algorithm for seeking peaks in the three LIGO-Virgo XWV spectra 
which are {\it  consistent} with the constraints 
\beq
\left\{
\begin{array}{l}
|t_{23}|=2\left|
T_{12}-T_{13}
\right|
\leq
c^{-1}
\left|
\vec{r}_{23}
\right|\\
|t_{13}|=2\left|
T_{23}-T_{12}
\right|
\leq
c^{-1}
\left|
\vec{r}_{13}
\right|\\
|t_{12}|=2\left|
T_{23}-T_{13}
\right|
\leq
c^{-1}
\left|
\vec{r}_{12}
\right|
\end{array}
\right.
\eeq 
expressing the obvious requirements that the wavefront propagation delay between two detectors
cannot exceed the limiting value corresponding to propagation along the line-of-sight direction
between the detectors, can be now formulated. The algorithm uses  the three (discrete, noisy) XWV spectra 
to construct a grid  in the time delay plane $(t_{13},t_{23})$, and assign different {\it levels}  $R$ to
its nodes:\\
\\
{\tt$~~~$initialize all time-delay grid node levels to zero}\\
{\tt$~~~~~$for all time-frequency pairs $(T_{12},f_{12})$ in $W_{12}$}\\
{\tt$~~~~~~~$ for all time-frequency pairs $(T_{13},f_{13})$ in $W_{13}$ such that:}\\
{\tt$~~~~~~~~~~~2|T_{12}-T_{13}| \leq c^{-1}|\vec{r}_{23}| \mbox{ and } f_{13}=f_{12} $}\\
{\tt$~~~~~~~~~~~~~~~$ for all time-frequency pairs $(T_{23},f_{23})$ in $W_{23}$ such that:}\\
{\tt$~~~~~~~~~~~~~~~~~~~2|T_{12}-T_{23}| \leq c^{-1}|\vec{r}_{13}| \mbox{ and } f_{23}=f_{12} $}\\
{\tt$~~~~~~~~~~~~~~~~~~~~~~~$accumulate level $R =R+|W_{12}(T_{12},f_{12})W_{13}(T_{13},f_{13})W_{23}(T_{23},f_{23})|$}\\
{\tt$~~~~~~~~~~~~~~~~~~~~~~~$at grid node $\{t_{13}=2(T_{12}-T_{23}),t_{23}=2(T_{12}-T_{13})\}$}\\
{\tt$~~~~~~~~~~~~~~~$ end for}\\
{\tt$~~~~~~~$ end for}\\
{\tt$~~~~~$end for}.\\
\\
A {\it candidate} direction of arrival is obtained by taking 
the highest-level grid-node in the $(t_{13},t_{23})$ plane subset defined by the further
constraint\footnote{
As shown in Sect. 4, the bound in (\ref{eq:strip}) can be made slightly tighter.}.
%
\beq
\left|
t_{12}
\right|
=
\left|
t_{13}-t_{23}
\right|
\leq
c^{-1}
\left|
\vec{r}_{12}
\right|
\label{eq:strip}
\eeq
The highest level in the grid can be used both as an {\it estimator} of the DOA, and as a {\it detection statistic},
whose performances will be discussed in Sect. 5.1.\\
Note that the proposed algorithm is {\it coherent}, 
in the sense that it produces a {\it single} detection statistic
by combining the data from {\it all} detectors in the network. 
It also inherits the typical features of {\it coincident} tests: 
the outermost loop enforces frequency-consistency, 
while the two inner loops enforce time-delay {\it admissibility}.\\
Note also that the XWV spectra will display sensible peaks only if the  waveform gathered by the different
detectors are {\it consistent in shape}. This suggests that the algorithm will be robust
against (independent) instrumental disturbances, as further illustrated in Sect. 6. \\
%
\subsection{LIGO-Virgo Network Directional Response under XWV Based Algorithm }
%
In the absence of noise, the above algorithm  
will produce a peak  in the time-delay grid
whenever a GWB is observed by the LIGO Virgo network.  
The peak will be located at a node whose time-delay coordinates
correspond to the DOA $(\vartheta_s,\varphi_s)$.
This peak will be well localized  provided  the duration of the transient signal is substantially shorter than
the minimum graviton flight-time between detectors.\\
In view of the bilinear nature of the XWV, it can be argued that the peak height  will be proportional to the squared
product of the three detectors' pattern functions along that direction.
This quantity, normalized to its maximum, and denoted henceforth as $\overline{\Phi}(\vartheta,\varphi)$  
describes the directional response of the proposed GWB detector/DOA estimator,
and is plotted in Figure 3 for the LIGO-Virgo network, for circularly polarized GWs.
We checked numerically that the expected (normalized) levels of the time-delay grid peaks, 
reproduce those computed from the function $\overline{\Phi}$ for each DOA in a $\vartheta,\varphi$ grid of
$50 \times 100$ points (using $10^3$ noise realization for each DOA). \\
The quantity:
\beq
\frac{\Omega[\overline{\Phi}_{min}]}{4\pi}=
\frac{1}{4\pi}\int_{\overline{\Phi}(\vartheta,\varphi) > \overline{\Phi}_{min}} \sin\vartheta d\vartheta d\varphi
\eeq
expresses the {\it fraction} of the (unit) celestial sphere where the (normalized) directional response of the
proposed detector/estimator exceeds the threshold value $\overline{\Phi}_{min}$, and is displayed in Figure 4.\\ 
It is seen, e.g., that roughly $50\%$ of the celestial sphere is covered  with $\overline{\Phi}_{min} \geq .2$
by the LIGO-Virgo network, using the proposed algorithm.
%
\section{DOA Reconstruction Uncertainties}
%
Uncertainties in the DOA reconstruction stem from a twofold origin: the discreteness of the time-delay grid,
due to the discrete implementation of the XWV spectra (finite time resolution), 
and the additive noise in the data (see discussion in Sect. 5.2).\\
In order to translate the effect of systematic and statistical errors in the estimated delays
into uncertainty ranges in the estimated DOAs, it is expedient to introduce the projection 
which maps the DOA  polar angles ($\vartheta_s,\varphi_s)$ into a point $(x_s, y_s)$ 
of the disc (with center  $O$ and radius $R$)  going through the detector, viz.\footnote{
We recall that the $x$-axis goes through detector-1 (LIGO-Livingston).}:
%
\beq
x_s=R\sin\vartheta_s\cos\varphi_s,\mbox{ }
y_s=R\sin\vartheta_s\sin\varphi_s.
\eeq
The formula which relates the $(x_s,y_s)$ projection to the arrival-time delays is obtained
from eq. (\ref{eq:delays}),
\beq
\left\{
\begin{array}{l}
\displaystyle{
x_s= c \frac{(t_{13}-t_{23}) \sin \varphi_3 -t_{13} \sin \varphi_2
     }
     { - \sin (\varphi_2-\varphi_3) 
               + \sin \varphi_2 -\sin \varphi_3
     }}\\
\\
\displaystyle{
y_s=c \frac{ t_{13} \cos \varphi_2 
      - (t_{13} - t_{23}) \cos \varphi_3  - t_{23}
     }
    {-\sin (\varphi_2-\varphi_3)
              +\sin \varphi_2 - \sin \varphi_3           
    }}
\end{array}
\right.
\label{eq:lintra}
\eeq
Equation (\ref{eq:lintra}) is a {\it linear} transformation, 
relating not only the coordinates ($x_s, y_s$)  to the delays $(t_{13}, t_{23})$,
but also the {\it uncertainties} $\delta x_s$, $\delta y_s$
to the delay errors $\delta t_{13}$ and $\delta t_{23}$.
Thus, under the simplest assumption where these latter are independent
and identically distributed, the uncertainty region in the  ($t_{13}, t_{23}$) plane is a circle, 
and the corresponding uncertainty region in the $(x_s,y_s)$ plane is an {\it ellipse}. 
Notably, the {\it shape} of this latter is translation-invariant across the circle $x_s^2+y_s^2 \leq R^2$,
i.e., DOA independent.\\
The ratio between the uncertainty areas in the  $(x_s,y_s)$ and ($t_{13}, t_{23}$) planes 
is given by the Jacobian of the transformation (\ref{eq:lintra}),
viz.:
\beq
J=c^2
\frac{\sin(\varphi_3-\varphi_2)}{\sin\varphi_3-\sin\varphi_2+\sin(\varphi_2-\varphi_3)}
\label{eq:jacob}
\eeq
which is also DOA-independent.\\
By back-projecting the uncertainty ellipse onto the  sphere of radius $R$ centered at $O$, 
we obtain a DOA-dependent uncertainty region.
This is illustrated in Figure 5,  for a few representative cases.
The ratio between the area of the uncertainty region on the celestial sphere, 
and the area of the uncertainty ellipse in the $(x_s,y_s)$ plane is displayed in Figures 6a and 6b
as a function of $\varphi_s$, for various values of $\vartheta_s$. 
For $\vartheta_s \sim 0$, this ratio is close to unity, whatever $\varphi_s$. 
On the other hand  as $\vartheta_s\rightarrow \pi/2$, the ratio blows up, 
and its dependendency on $\varphi_s$ becomes more and more evident.
Such a behaviour had been already noted in, e.g., \cite{Cavalier_06}, \cite{Fairh_09}.
%
\section{Numerical Experiments}
%
In order to check the performance of the proposed algorithm, we run a series of Monte Carlo simulations.
The simulations use  time-discretized GWBs and glitches 
injected into  white (independent) random Gaussian sequences, to represent the three interferometer data.
In the case of GWBs, the delays are chosen according to the assumed source location.\\ 
Our XWV engine uses  data chunks $2048$ time-samples wide 
to produce  a $1024 \times 1024$  time-frequency nodes XWV transform, 
using Pei-Yang fast algorithm \cite{PeiYang}. The sampling frequency is $4\mbox{  }KHz$. 
The data are accordingly decomposed into half-overlapping chunks $2048$ time-samples wide
(we use a plain rectangular windowing function), in order to use a fixed number of time samples
to compute each time-frequency samples.
The resulting discrete XWV spectrum spans the time range between samples $\#513$ and $\#1536$,
and the frequency range between $0$ and $1000\mbox{ }Hz$\footnote{
Note that when using analytic signals for computing discrete versions of the XWVT, the minimum sampling rate 
must be {\it twice} the Shannon rate \cite{XWV2}}.
%
As already mentioned, the XWV values are equalized so that in the absence of signals their first and
second moment  are $0$ and $1$, respectively.\\
Now, even in the absence of a signal, the  levels produced by our algorithm 
in the time-delay plane  $(t_{13},t_{23})$  grid-nodes will be {\it non}-uniform,
due to the {\it different} number of (noisy) time-frequency XWV values mapped into each node. 
The average and standard deviation in the $(t_{13}, t_{23})$ plane for pure-noise (stationary, Gaussian)
data are shown in Figure 7.
We accordingly {\it equalize} the  levels, 
by subtracting the above average, and dividing the result by the above standard deviation, so that 
in the absence of signals, the grid-node levels in the time-delay plane will
have zero average and unit variance.\\  
For each injected waveform, we generated $10^4$ different noise realizations, to test
the statistical properties of the proposed algorithm, both as a detector and as a DOA estimator.
The waveforms were parameterized by their {\it intrinsic} signal to noise ratio (SNR), defined by
\beq
\delta_h=\frac{h_{rss}}{N}=\frac{
\left\{
\int
\left[
h_+^2(t)+h_\times^2(t)
\right]dt
\right\}^{1/2}
}{N},
\label{eq:deltah}
\eeq
$N$ being the (two-sided) power spectral density of the stationary white(ned) Gaussian noise component,
assumed for simplicity the same in all detectors (the effect of glitches will be discussed in Sect. 6).\\
The results of our simulations are summarized below.
%
\subsection{Detection Performance}
%
The performance of our algorithm as a detector are illustrated in Figures 8 and 9.
The detection statistic is the level of the highest peak in the time-delay grid.
Figure 8a and 8b display the false alarm (continuous line) and false dismissal probabilities (the dashed
lines, corresponding to different values of the intrinsic SNR ($\delta_h$) as functions
of the detection threshold $\gamma$, for  DOAs corresponding to 
the maximum ($\vartheta=0.705rad$, $\varphi=5.073rad$) 
and the minimum ($\vartheta=0.800rad$, $\varphi=1.100rad$)
of the network pattern function in Fig. 3.
Figures 9a and 9b show the receiver operating characteristics, i.e., the detection probability
vs the false alarm probability, for fixed values of the intrinsic signal to noise ratio, $\delta_h$,
for a DOA corresponding to the maximum of the network pattern function in Fig. 3.
%
\subsection{DOA Estimation Performance}
%
As already mentioned, the finite time resolution implies that the estimated delays are affected by a {\it systematic}
uncertainty which can be twice the XWV time-step $\delta t$.
The noise in the data entails that estimated delays  {\it spread} around the {\it actual} delays
in a signal-to-noise dependent way. This is illustrated in Figures 10 and 11. 
The estimate is always unbiased, whenever the signals are shorter than the minimum
graviton flight time between the detectors.
Figure 10 displays the standard deviation of the estimated delays (average between the two)
as a function of the intrinsic SNR, for DOAs corresponding to 
the maximum ($\vartheta=0.705rad$, $\varphi=5.073rad$) 
and the minimum ($\vartheta=0.800rad$, $\varphi=1.100rad$)
of the network pattern function in Fig. 3. In a log-log scale,  both curves show the same slope,
corresponding to an exponent $\approx -1.4$.
Figure 11a displays the empirical distribution of the estimated delays for $10^4$ different noise
realizations, for a DOA corresponding to the maximum of the network pattern function in Fig. 3,
for two different values  of the intrinsic signal to noise ratio.\\
It is interesting to compare Fig. 11a to Fig. 11b, where a standard correlation-based 
time-delay estimator \cite{Marple}  has been used to retrieve the two propagation delays.
Our XWV-based  estimator is seen to offer distinctly better performances.
%
\subsection{Realistic Waveforms}
%
As anticipated in Section 2, the XWV transform peak localization property
holds not only  for SG waveforms, but for general transient waveforms.
This is further illustrated in Figures 12 to 15.\\
Figure 12 (top) shows two copies of a typical supernova GWBs, 
belonging to the family computed by Dimmelmaier and co-workers \cite{Dimmel}, 
with a time shift of $82$ time samples (corresponding to $20.5$ $ms$, at our sampling frequency), 
together with their XWV transform. 
The XWV is identical to the Wigner transform
of the GWB waveform, except for the time-shift given by eq. (\ref{eq:WVpeak}).
Accordingly, the XWV peak in Fig. 12 (bottom) is localized at the midpoint between the
peak times of the two waveforms  in Fig. 12 (top).
Figure 13 (left) shows the  time-delay grid histogram when this waveform 
is emitted by a source located in the direction
of maximum network sensititivity in Fig. 3, for $\delta_h=100$. 
It can be seen that, not unexpectedly,
the localization properties in the delay plane are even better than for Gaussian waveforms,
in view of the larger time-bandwith figure of the Dimmelmaier waveform.\\
Figures 14 and 15 are similar to Figs. 12 and 13, 
except that the waveform here is that of a typical binary merger \cite{merger}.
\\
%
\section{Glitch Rejection}
\label{sec:Glitches}
%
By construction the proposed detection/localization algorithm should be robust against spurious transients
of environmental/instrumental origin (glitches).\\
We may expect that glitch induced false detection may occur
only in the (unlikely) case where {\it each} detector shows a glitch in the analysis window,
such that the mutual delays are consistent with an acceptable DOA,
{\it and} the (indipendent) glitch waveforms are {\it consistent} in shape.\\
In order to illustrate these features, we consider first the no-GWB case  where  a  glitch  occurs 
in the data of {\it each} of the three interferometers, 
the three glitches being  {\it different},  but with delays consistent with an {\it admissible}  DOA.\\ 
To this end, we used a set of $N=7$  visually different waveforms,  shown in Figure 16 (top),
from the catalogue of "typical" LIGO glitches compiled by P. Saulson \cite{glitch_zoo}. 
All glitches in the set were scaled to unit norm, and time-shifted so as to bring their envelope peaks to coincidence.  
The (normalized) pairwise correlation coefficient of the selected glitches, 
which provides some quantitative measure of their (dis)-similarity, 
does not exceed $0.62$, with mean and median values of $0.195$ and $0.105$, respectively. 
The correlation coefficient histogram is shown in Figure 16 (bottom left).\\  
From the above glitch set, we formed (all)  $35$ triplets of {\it different} waveforms, 
and computed the related X-Wigner transforms and the peak-levels in the time-delay grid produced by our algorithm.  
For each of glitch-triplet  $(g_1,g_2,g_3)$  we also computed the geometric mean 
of the time-delay grid peak-levels  for the three cases 
where the data from {\it all} interferometers contain the {\it same} waveform \footnote{
These correspond to a source {\it equidistant} from all detectors radiating the waveform $g_i$,
with all detectors exhibiting the {\it same} response in the source direction. 
The last assumption is unrealistic, but is irrelevant for the present purpose).}
$g_i$,  $i=1,2,3$. 
This quantity was used to re-scale the time-delay grid peak-level for that glitch-triplet.\\
The histogram of the rescaled time-delay grid peak-levels for the $35$ different glitch triplets considered is shown in Figure 16 (bottom right).  
The largest (scaled) peak level was $0.51$, with a median value of $0.12$ and a mean of $0.17$.\\
These, admittedly limited, results illustrate the waveform consistency test capabilities of the proposed algorithm.\\ 
\\
We next consider the case where GWBs {\it and} glitches co-exist in the data. 
In these further simulations we used SG glitches and GWBs, 
with glitch parameters (center frequency, peak position and carrier frequency) generated randomly and independently in each detector. 
The pertinent results are illustrated in Figures 17 to 20. 
These figures show the noisy waveforms (left column), 
the density maps of the XWV transforms (mid column), 
and the (normalized) level map in the $(t_{12},t_{13})$  time-delay grid.\\
In Fig. 17 we consider the simplest case where the GWB data in a single detector (H1 in this case) 
are corrupted by a single glitch in the analysis window.
The GWB  signal to noise ratios are $22.16$ (L1), $23.13$ (H1) and $31.85$ (V), 
for a (circularly polarized) source with $\delta_h=50$ at $\vartheta=2.58 rad$ and $\varphi=2.71 rad$.
The glitch signal to noise ratio in H1 is $22.59$. 
The glitch shows up clearly in the H1 data, and produces evident artifacts in the H1-V and H1-L1  XWV spectra.
Nonetheless, its effect on the time-delay level map is almost negligible.\\
In Fig. 18 we consider the (unlikely) case where the data in {\it each} detector are corrupted by (single) glitches in the analysis window,
with SNR values of $11.5$ (L1), $12.00$ (H1) and $16.53$ (V).  None of the glitches has a significant overlap 
with the GWBs; nonetheless, they produce artifacts in all XWV transforms. Also in this case the effect of these artifacts on the
detection/localization properties is negligible, as seen from the time-delay level map.\\
Not unexpectedly, the localization performance deteriorates significantly in the rather extreme situation
where glitches {\it overlap} the GWBs in the data. 
When this happens in such a way that {\it true} peaks in the XWV transforms are no longer
{\it resolvable} from spurious ones, the time delay level map topography is substantially blurred
in the neighbourhood of the {\it true} delays, resulting into more or less severe localization errors.
This is illustrated  in Figs. 19 and 20.\\
In Fig. 19 we have a (single) GWB-overlapping glitch in H1 
with SNR=$16.28$.  In Fig. 20 each detector is affected by a GWB-overlapping glitch.
The glitch SNR values in Figs. 19 and 20 are the same as those in Figs. 17 and 18.
A sensible distortion in the  XWV transforms is observed, 
entailing a sensible error in the estimated delays.
%
\section{Conclusions}
%
We presented a simple, computationally light and fast algorithm for detecting short unmodeled GWB 
in a network of three interferometric GW detectors, 
and estimating the related DOA, based on XWV spectra. 
The algorithm is reasonably performant, and nicely robust against spurious transients (glitches) of instrumental origin
corrupting the (otherwise Gaussian) detectors noise floor.\\
It  does not provide waveform reconstruction; 
this latter, however can be accomplished in principle off-line, once the DOA has been estimated.\\
Generalization to larger networks, and other potentially interesting waveforms (e.g., chirps) is relatively straightforward.
Such extensions will be explored in a forthcoming paper.\\  
Based on the above preliminary results, we suggest that the proposed algorithm may be used  
as a quick-and-(not-so)-dirty  on-line data sieving tool.\\
A quantitative comparison  with existing GWB detection/DOA estimation
algorithms in terms of  efficiency and computational burden  will be the subject of future investigation.
%
\section*{Acknowledgements}
We thank the anonymous Referees for several useful suggestions and remarks.
%
\section*{References}


\pagebreak
\section*{Appendix - XWV Moments.}
\label{sect:appe}
%
Let 
\beq
W_{f,g}(n,m)=\sum^{L}_{k=-L}f(n+k)g^{*}(n-k)\exp[-4\imath N^{-1} \pi m k].
\eeq
the discrete version of the  XWV spectrum, where $n$ and $k$ 
are the discrete time and frequency index.
Note that, formally,
\beq
W_{f,g}(n_0,m)=DFT_x(2m)
\eeq
where $x=f(n_0+k)g^*(n_0-k)$. This shows that while
the time index $n$ spans the range $(-L, L)\cap\mathbb{N}$,
the frequency index $m$ spans the range $(-L/2,L/2)\cap\mathbb{N}$.\\
Let
\beq
\tilde{\nu} = \nu + \imath {\cal H}\nu 
\eeq
the analytic version of the background noise,
and denote as $\nu(k)$ and $\nu_H(k)$  
the (real-valued)  samples, of  the noise and its Hilbert transfom.\\
For zero average Gaussian white noise with (two sided) power spectral density $W_0$,
the spectral power density of analytic noise is 
\beq
S_{\nu_a,\nu_a}=4W_0 U[\theta(m)],
\eeq
where $U(\cdot)$ is Heaviside's function and $\theta=2\pi m/N \in (-\pi, \pi)$, 
$N=2L+1$ being the number of DFT frequency samples.\\
It is a simple task to show that the first moment of the XWV  is zero, in view of the
assumed independence of the (Gaussian) noises in different detectors.\\
We now compute the  second moment, viz.:
\[
\sigma^2(n,m)=E\left\{
W_{\nu_a,\mu_a}[n,\theta(m)]W^*_{\nu_a,\mu_a}[n,\theta(m)]
\right\}=
\]
\[
=\sum^{L}_{k=-L}\sum^{L}_{p=-L}
E[
\nu_a(n+k)\mu_a^{*}(n-k)
\nu_a^{*}(n+p)\mu_a(n-p)]
e^{-\imath 2  (k-p) \theta(m)}=
\]
\[
=\sum^{L}_{k=-L}\sum^{L}_{p=-L}
E[\nu_a(n+k)\nu_a^{*}(n+p)]
E[
\mu_a(n-p)\mu_a^{*}(n-k)]
e^{-\imath 2  (k-p) \theta(m)}=
\]
\beq
=\sum^{L}_{k=-L}\sum^{L}_{p=-L}
R^2_{\nu_a, \nu_a}(k-p)e^{-\imath 2 (k-p) \theta(m)}
\label{eq:app1}
\eeq
where $\nu_a$ and $\mu_a$ are built from independent, zero average, white(ned) Gaussian noises
pertinent to different detectors, but assumed as having the same power spectral density $W_0$,
and  $R_{\nu_a, \nu_a}(h-k)=E[\nu_a(h)\nu_a^{*}(k)]$ 
denotes the autocorrelation. function of analytic noise.
The inner summation in (\ref{eq:app1}) can be extended to $\infty$,
and the  Wiener Khinchin theorem can be invoked to prove that,
\[
\sigma^2(n,m)\approx
\sum^{L}_{k=-L}\sum^{\infty}_{p=-\infty}
R^2_{\nu_a, \nu_a}(k-p) e^{-\imath 2  (k-p) \theta(m)}=
\]
\[
=\sum^{L}_{k=-L}
S_{\nu_a, \nu_a}[2 \theta(m)] * S_{\nu_a, \nu_a}[2 \theta(m)]=\]
\beq
= 
16 W_0^2 \sum^{L}_{k=-L} \left| \frac{\theta}{\pi} \right|
=(2L+1) 16 W_0^2  \left| \frac{\theta}{\pi} \right|=
32W^2_0 |m|
\label{eq:app3}
\eeq
where $m \in [-L/2,L/2] \cap \mathbb{N}$.

\newpage

\begin{figure}
\centerline{\includegraphics[scale=0.5]{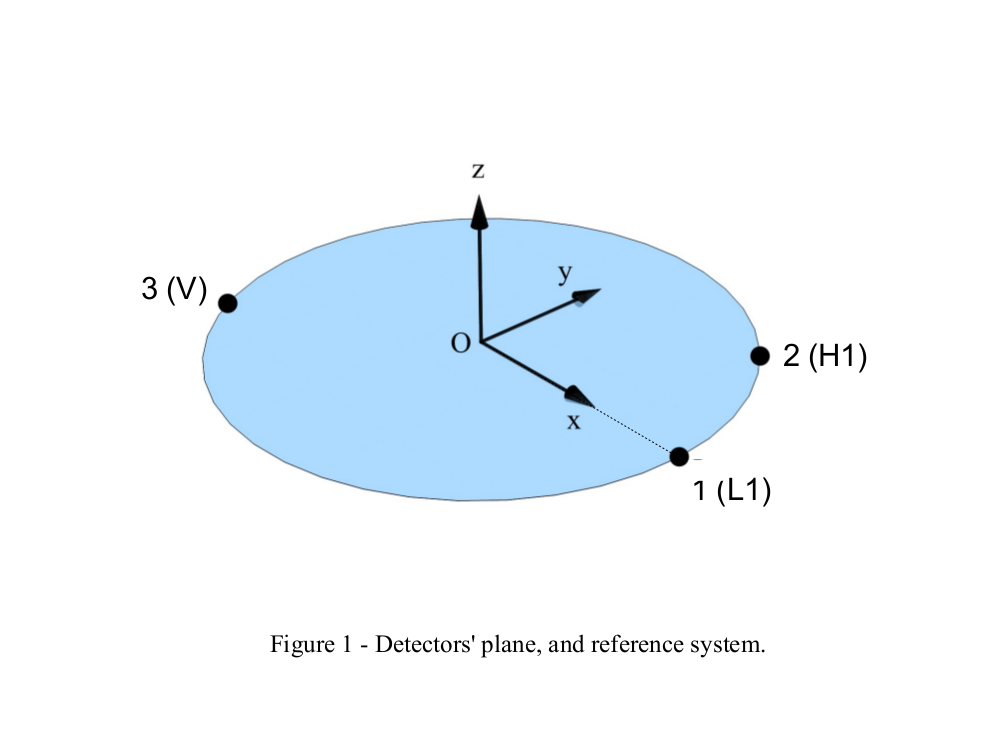}}
\centerline{\includegraphics[scale=0.5]{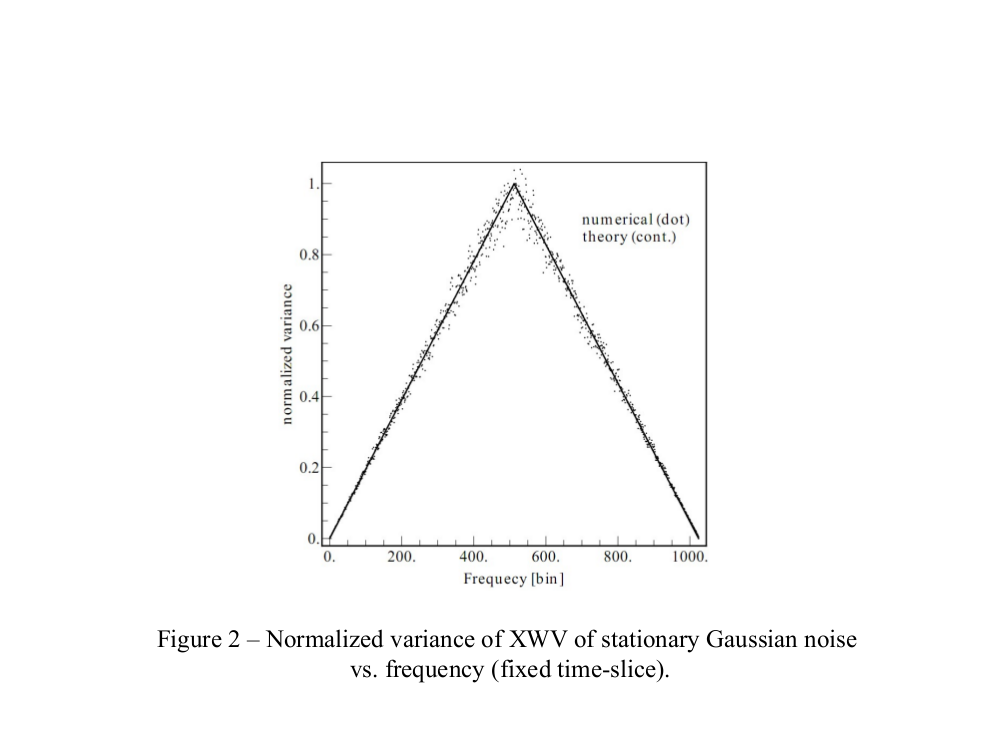}}
\end{figure}
\begin{figure}
\centerline{\includegraphics[scale=0.5]{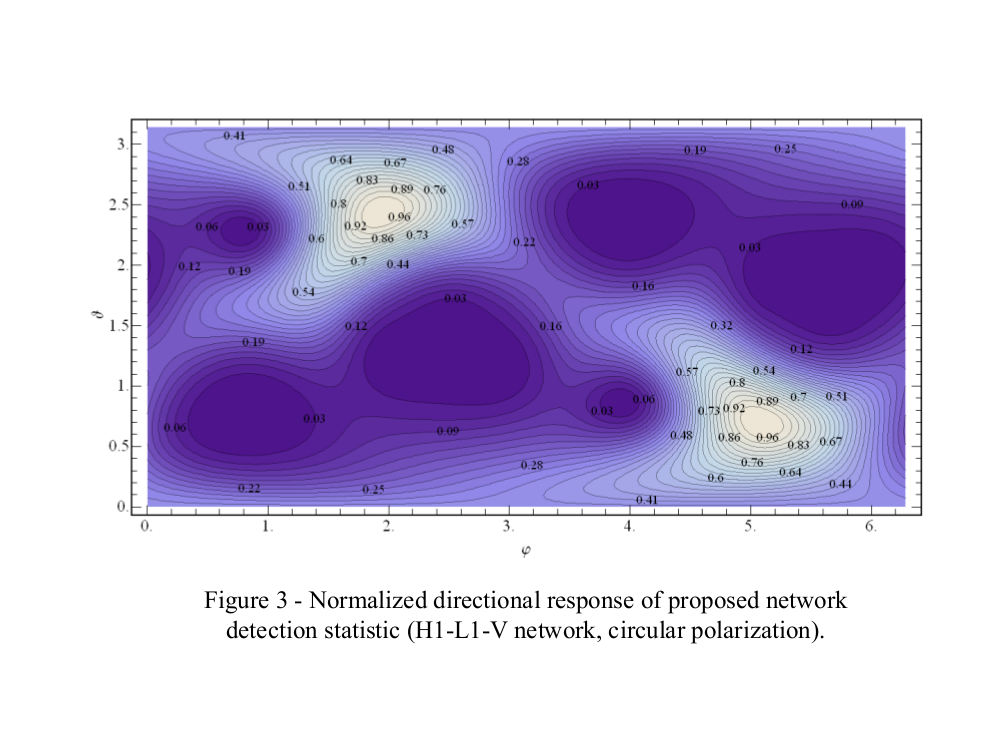}}
\centerline{\includegraphics[scale=0.5]{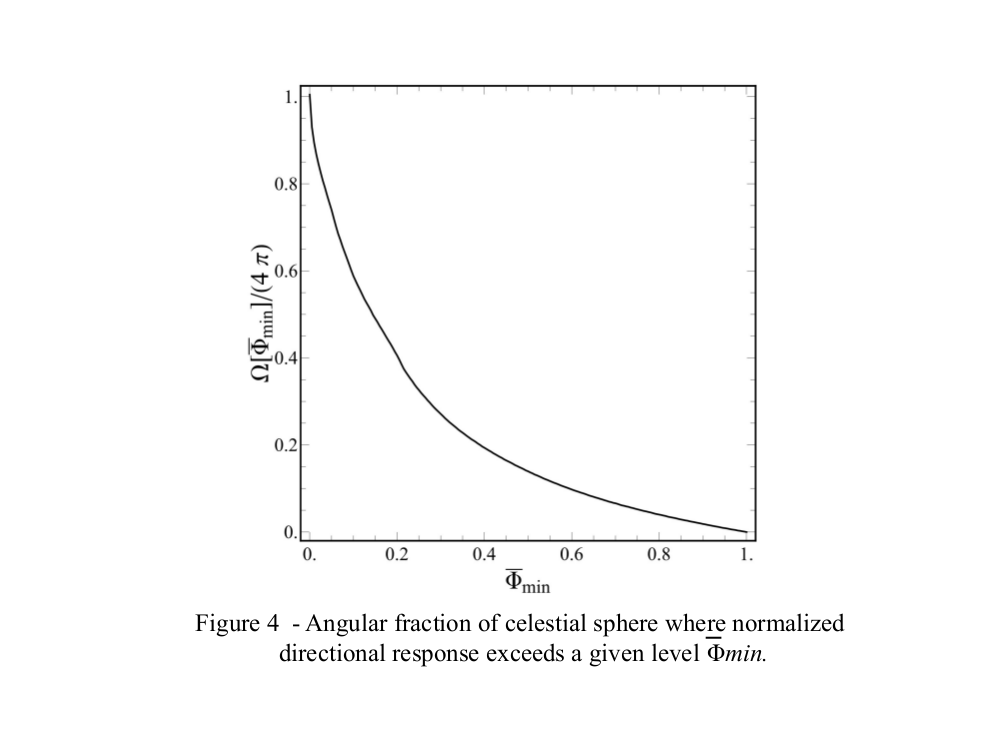}}
\end{figure}
\begin{figure}
\centerline{\includegraphics[scale=0.7]{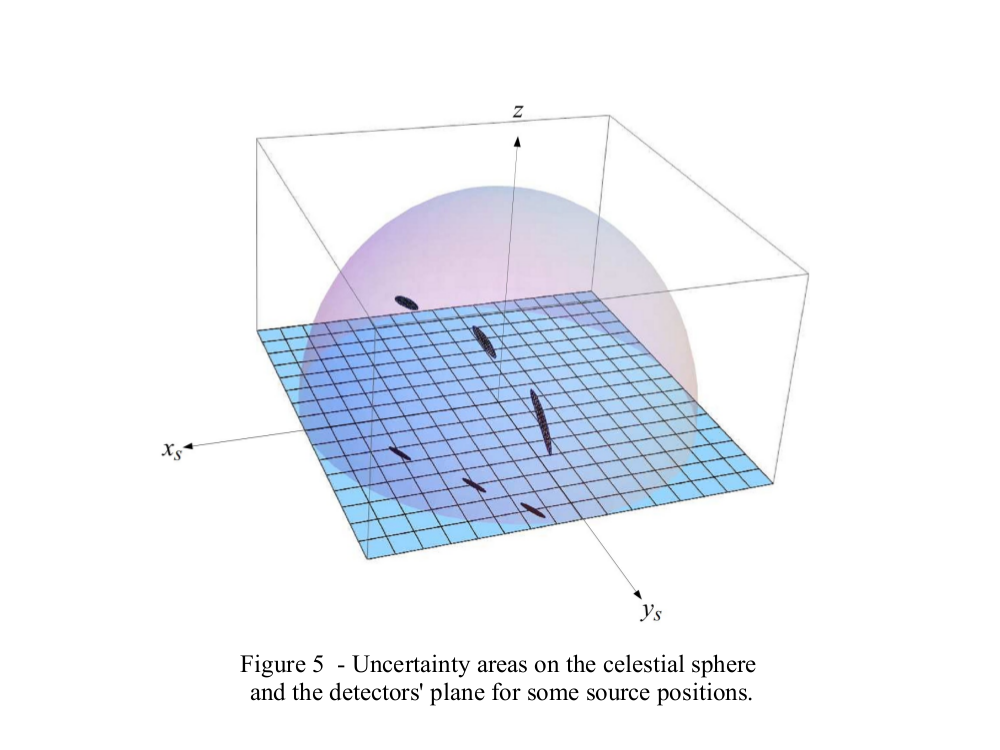}}
\end{figure}
\newpage
\begin{figure}
\centerline{\includegraphics[scale=0.5]{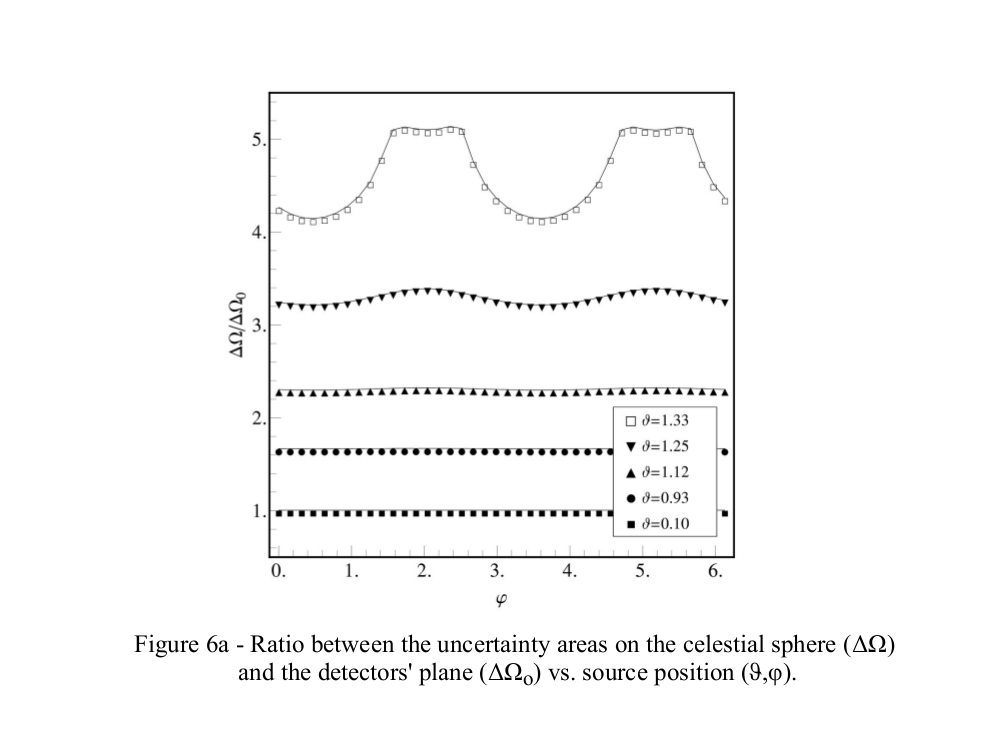}}
\centerline{\includegraphics[scale=0.5]{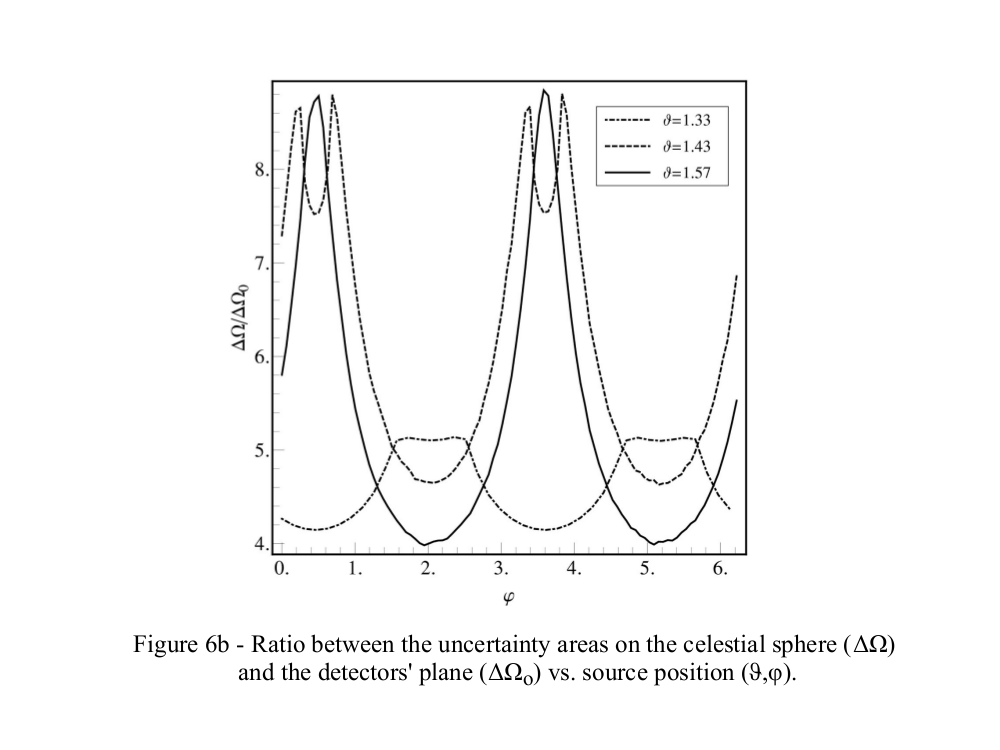}}
\end{figure}
\begin{figure}
\centerline{\includegraphics[scale=0.7]{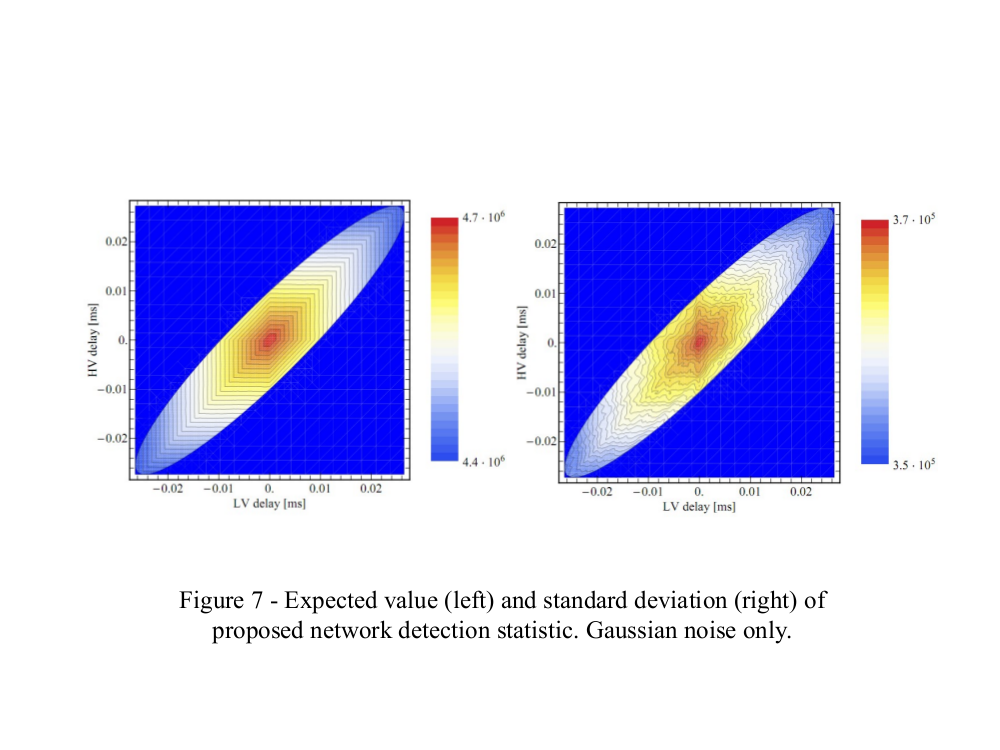}}
\end{figure}
\newpage
\begin{figure}
\centerline{\includegraphics[scale=0.5]{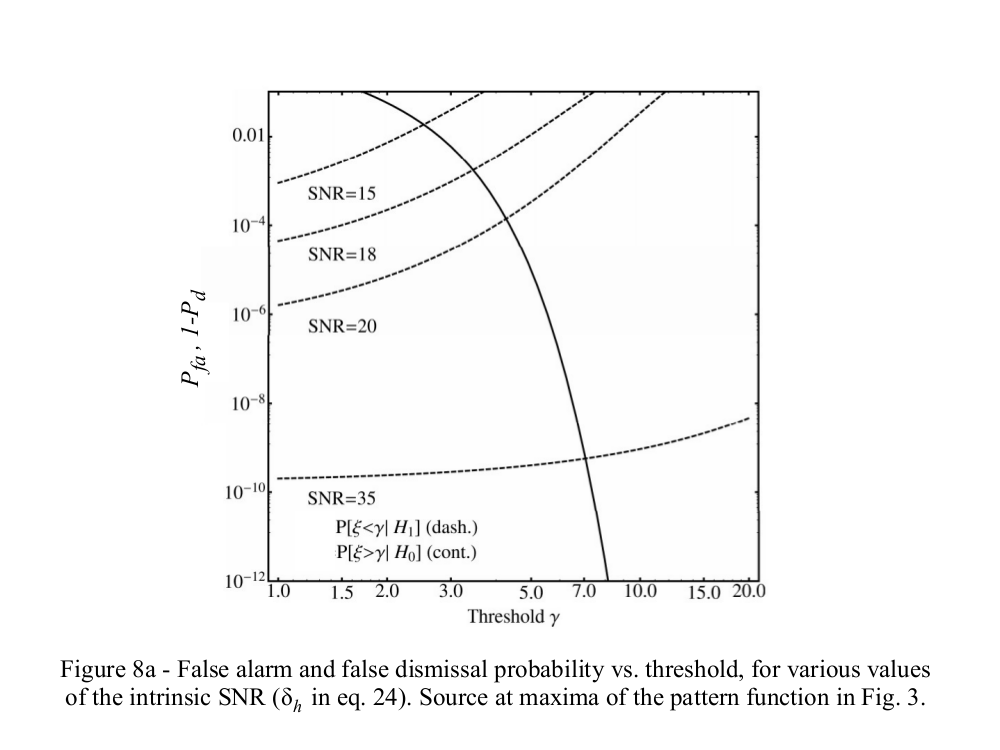}}
\centerline{\includegraphics[scale=0.5]{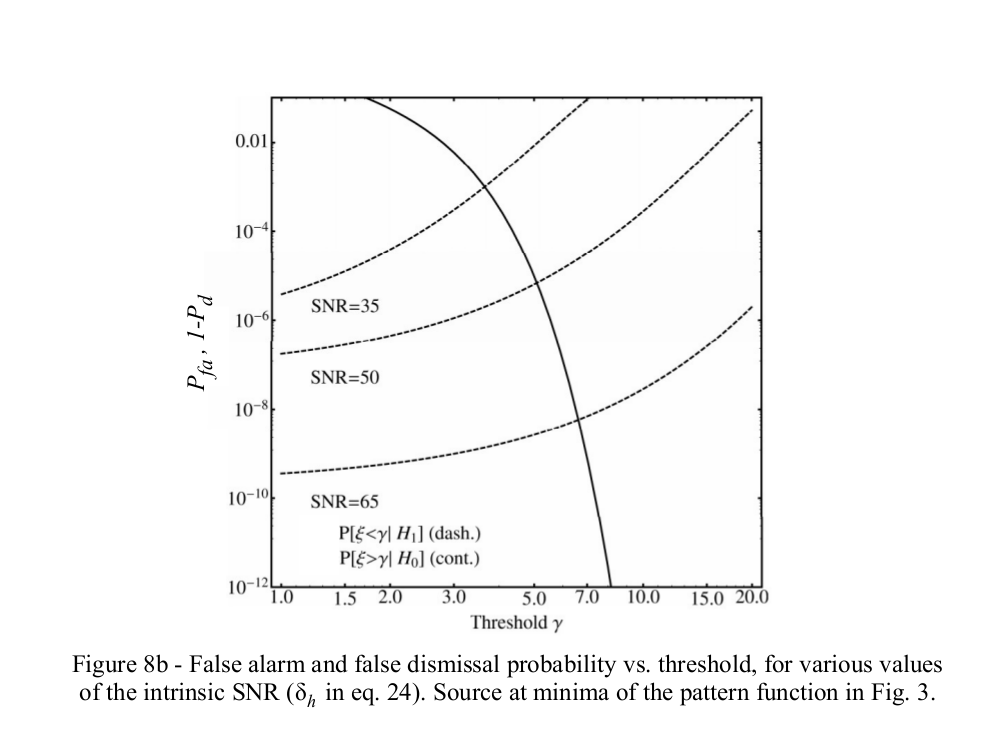}}
\end{figure}
\begin{figure}
\centerline{\includegraphics[scale=0.5]{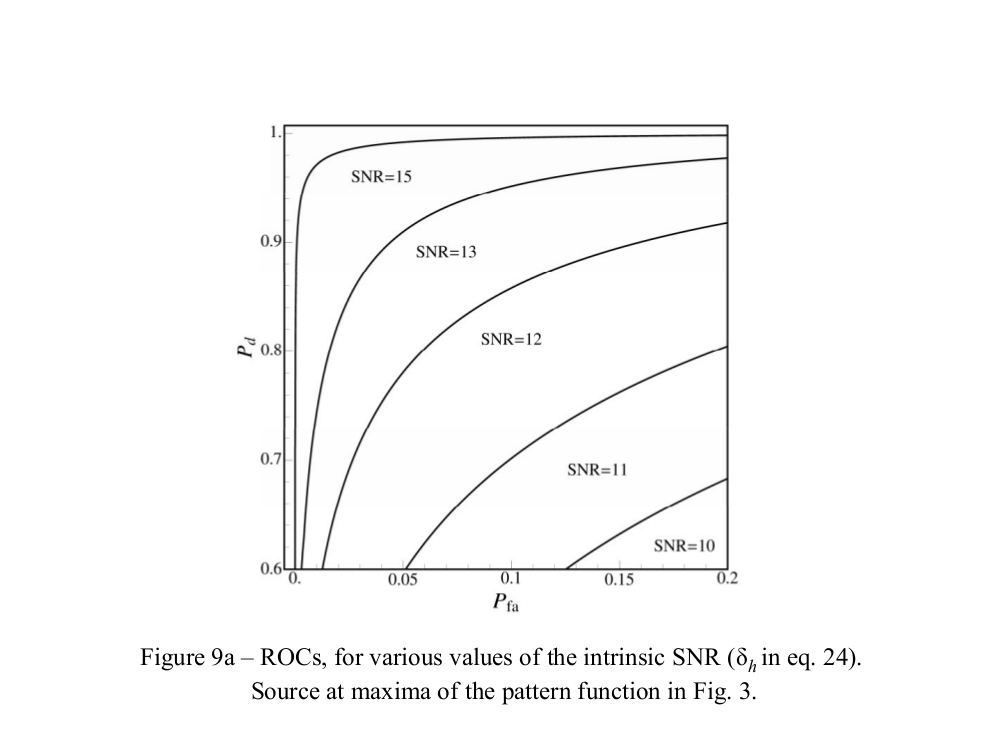}}
\centerline{\includegraphics[scale=0.5]{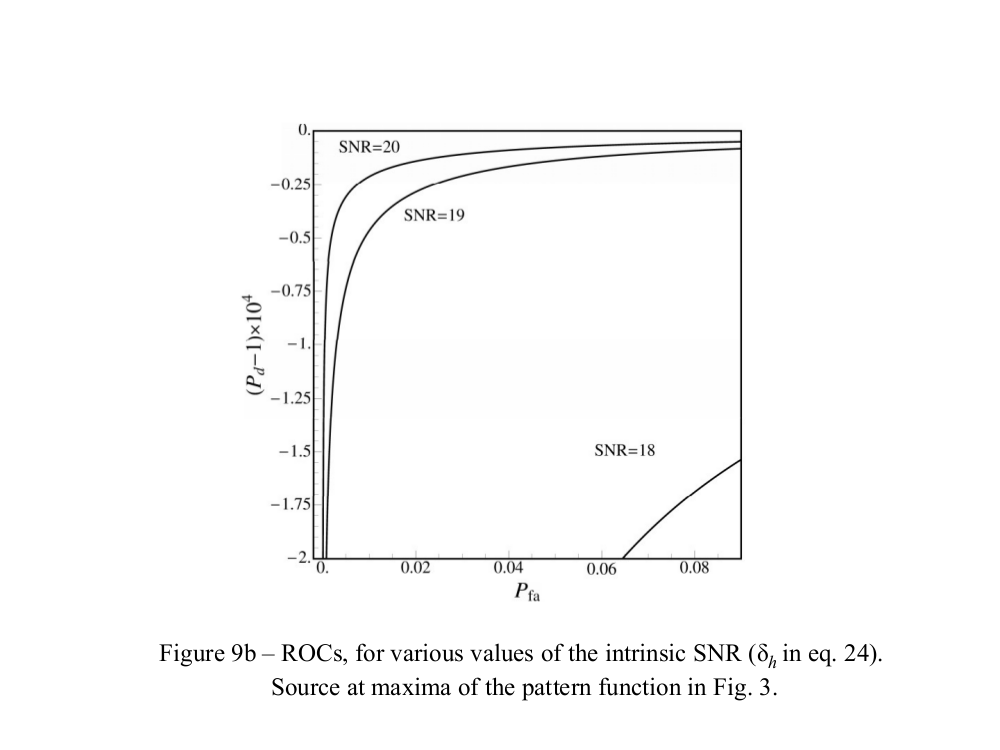}}
\end{figure}
\begin{figure}
\centerline{\includegraphics[scale=0.5]{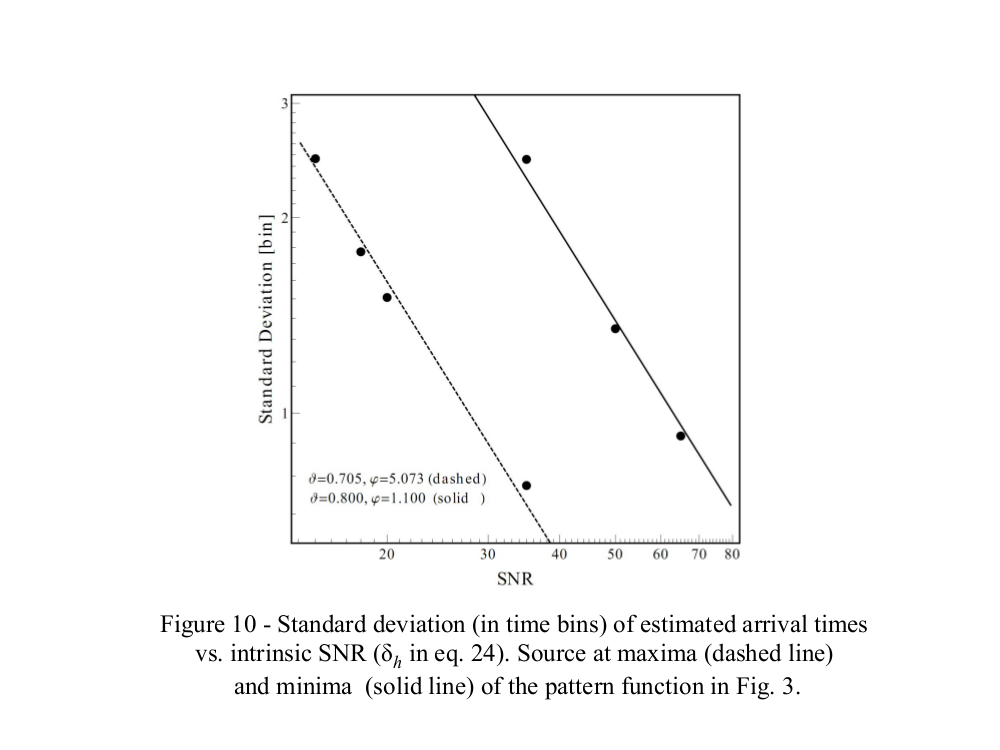}}
\end{figure}
\newpage
\begin{figure}
\centerline{\includegraphics[scale=0.5]{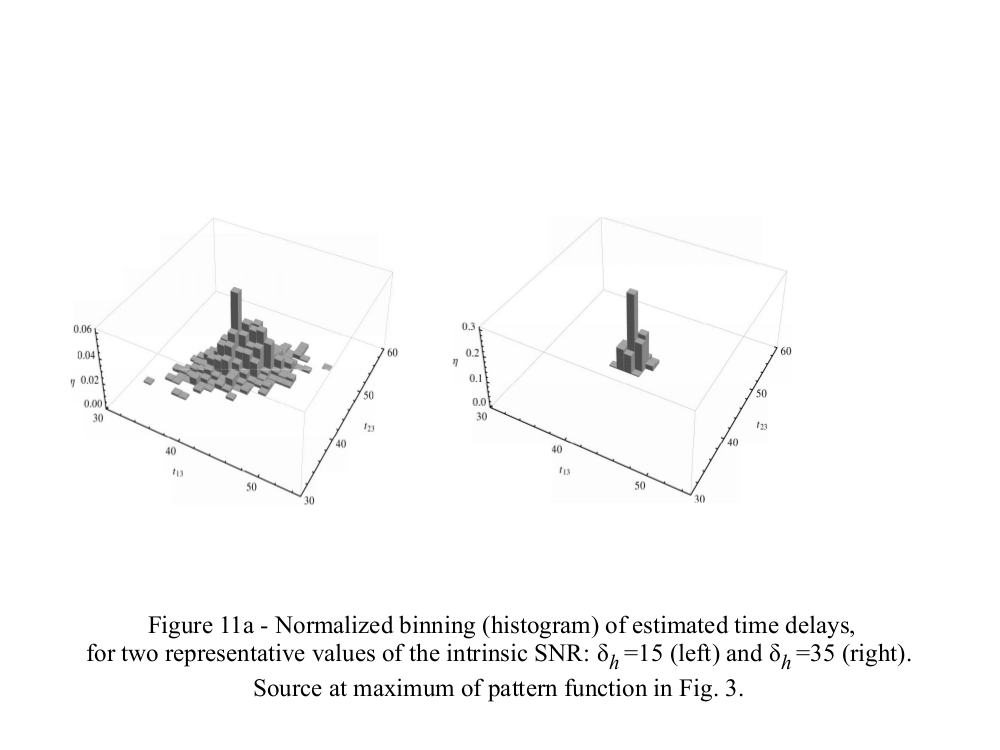}}
\centerline{\includegraphics[scale=0.5]{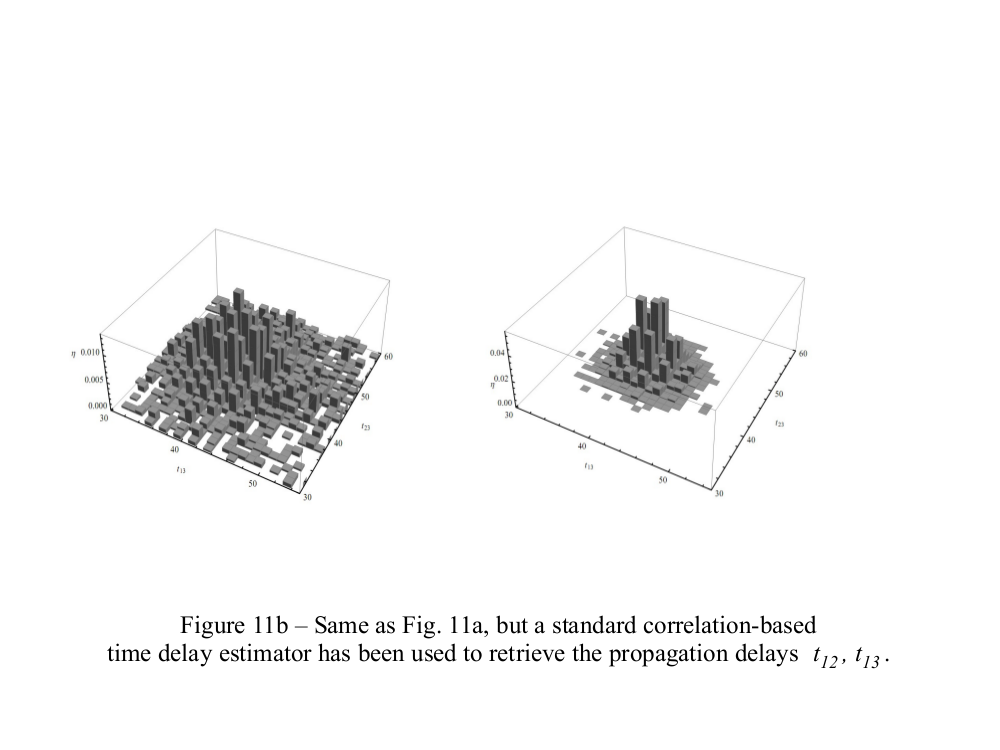}}
\end{figure}
\begin{figure}
\centerline{\includegraphics[scale=0.5]{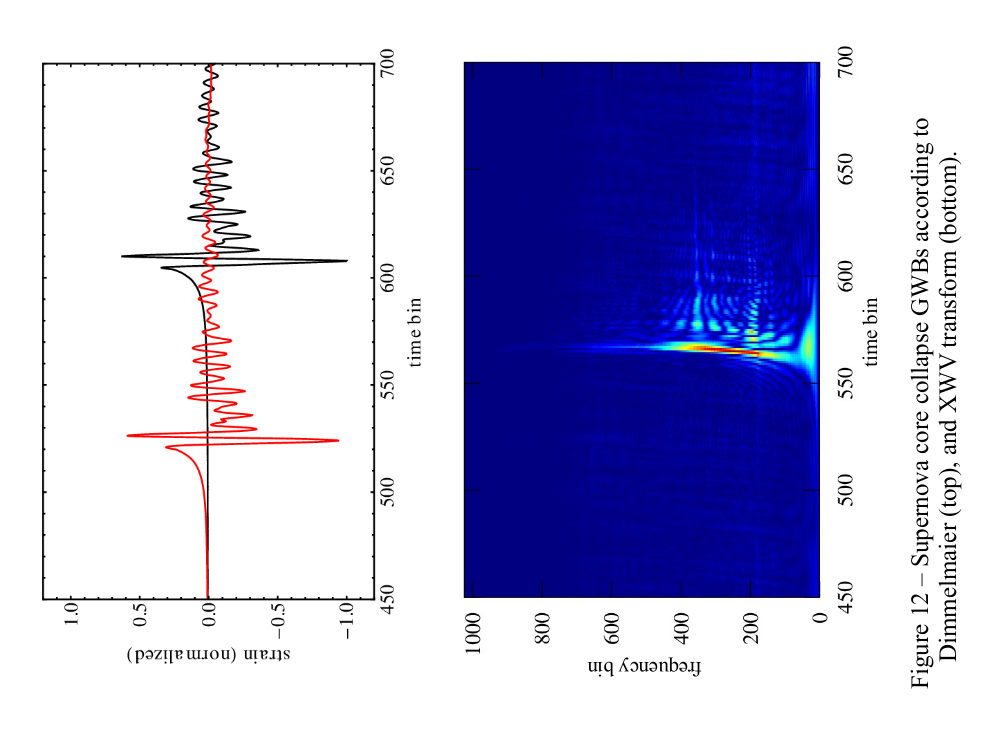}}
\centerline{\includegraphics[scale=0.5]{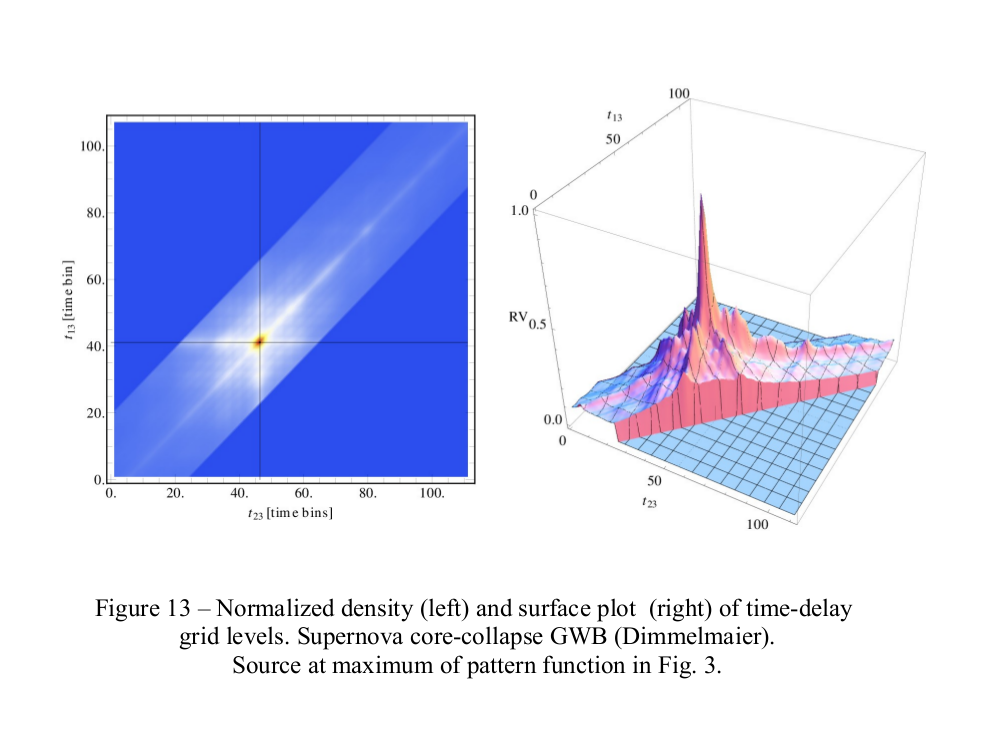}}
\end{figure}
\begin{figure}
\centerline{\includegraphics[scale=0.5]{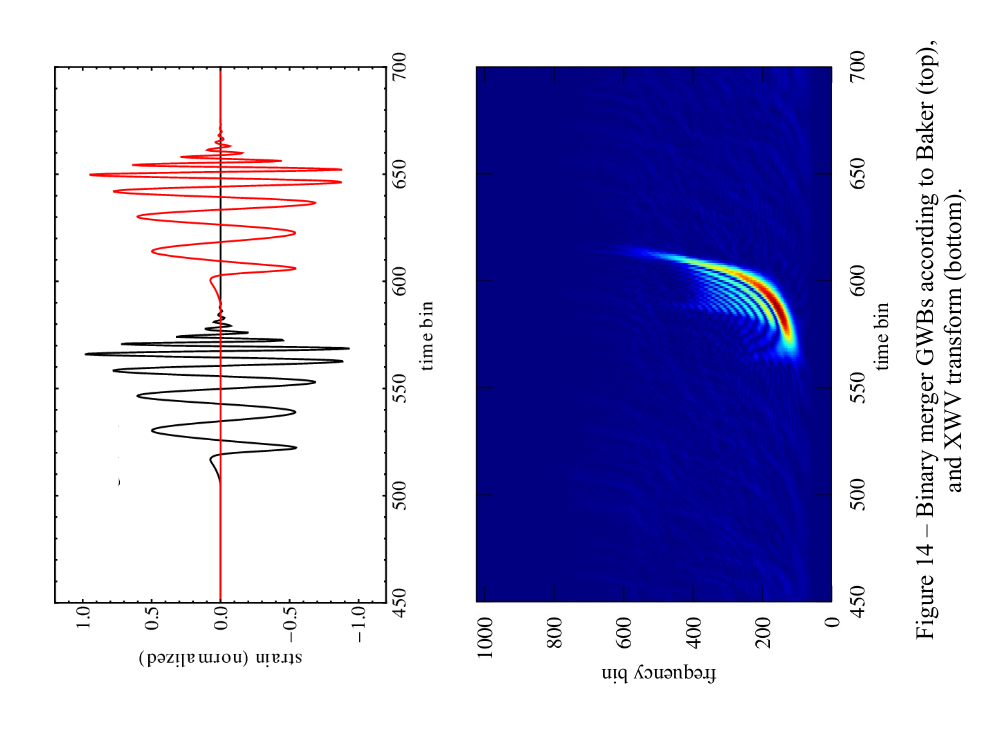}}
\centerline{\includegraphics[scale=0.5]{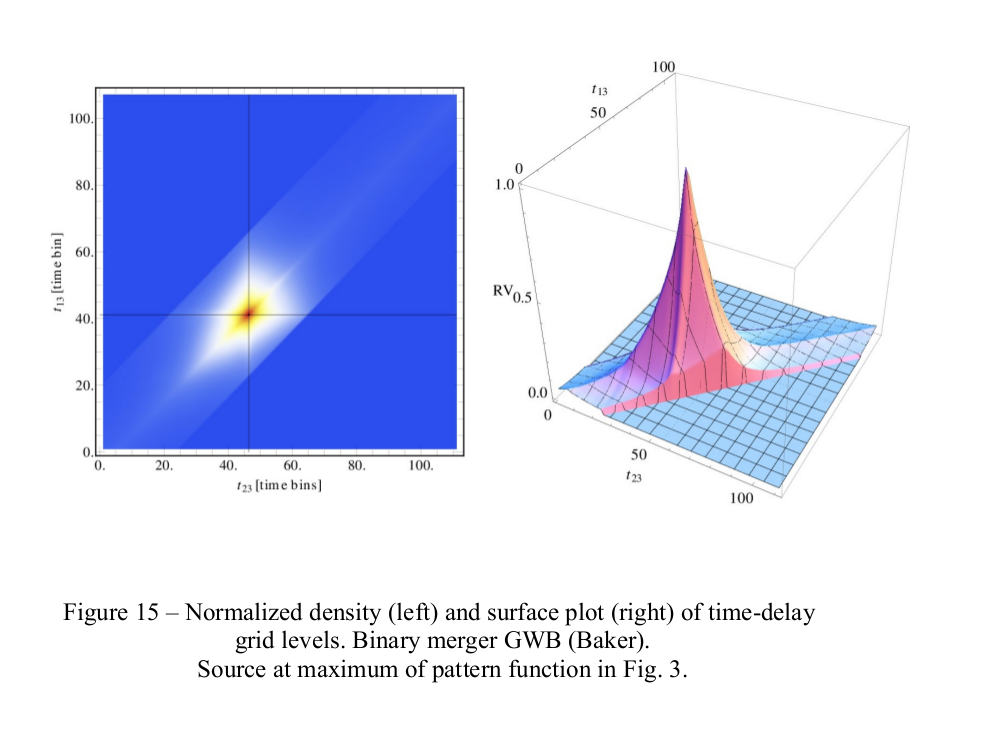}}
\end{figure}
\begin{figure}
\centerline{\includegraphics[scale=0.5]{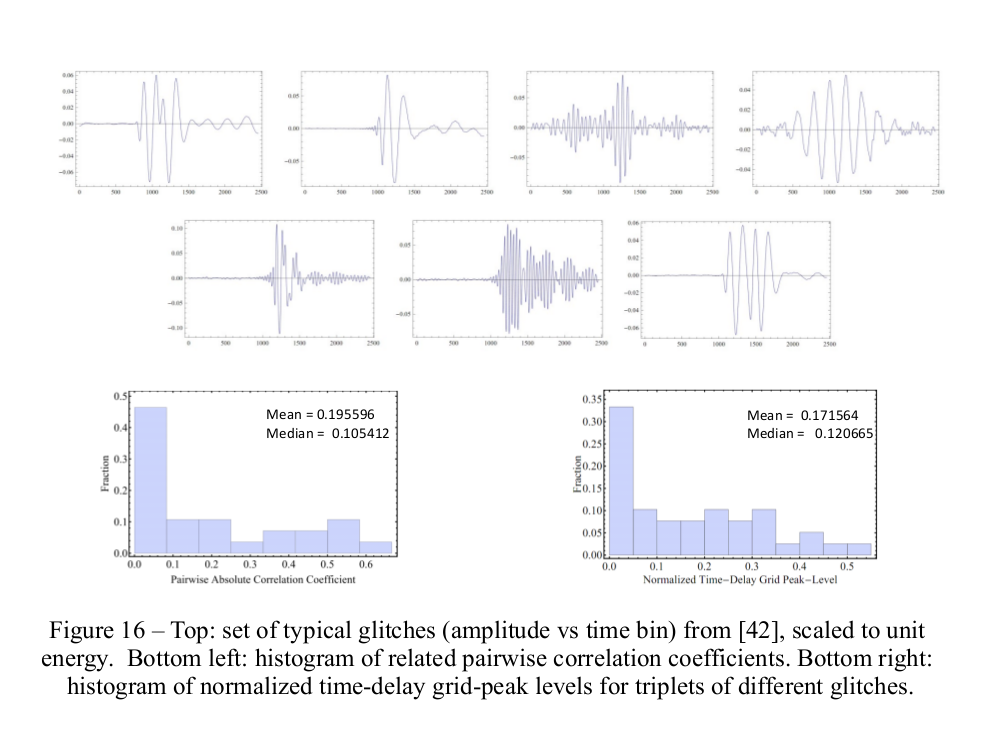}}
\end{figure}
\begin{figure}
\centerline{\includegraphics[scale=0.5]{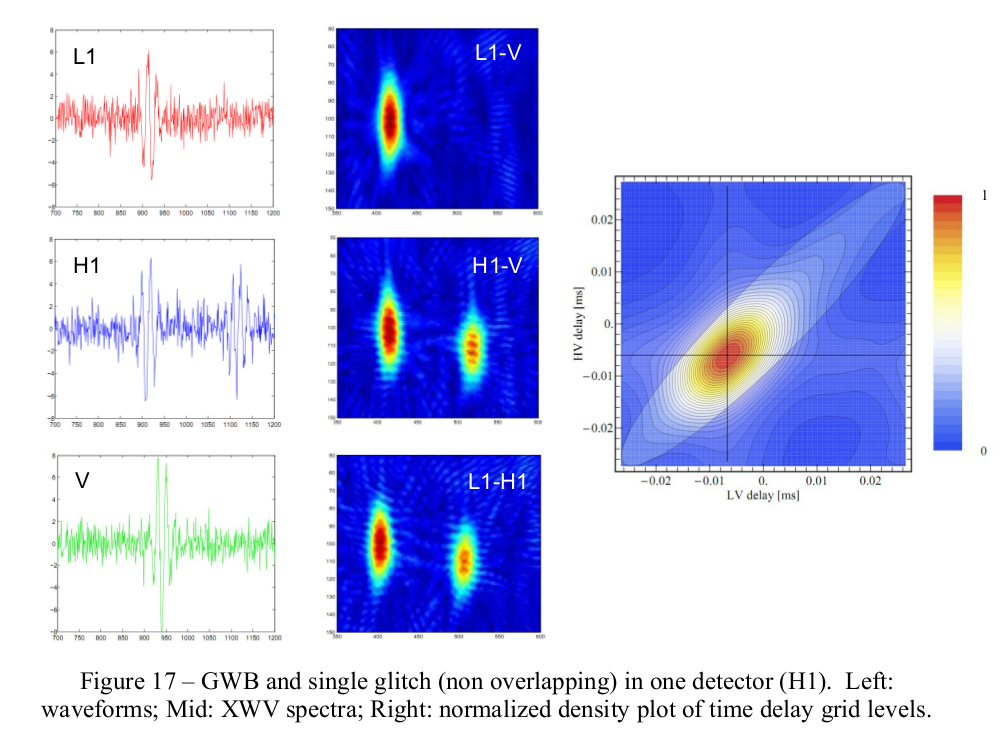}}
\end{figure}
\begin{figure}
\centerline{\includegraphics[scale=0.5]{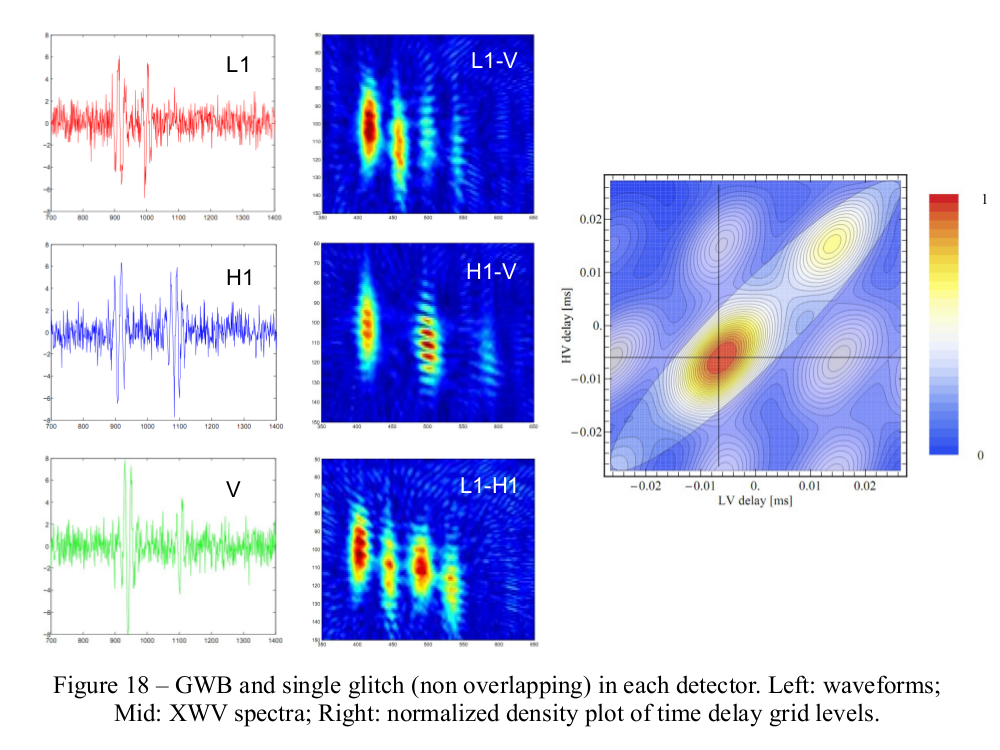}}
\end{figure}
\begin{figure}
\centerline{\includegraphics[scale=0.5]{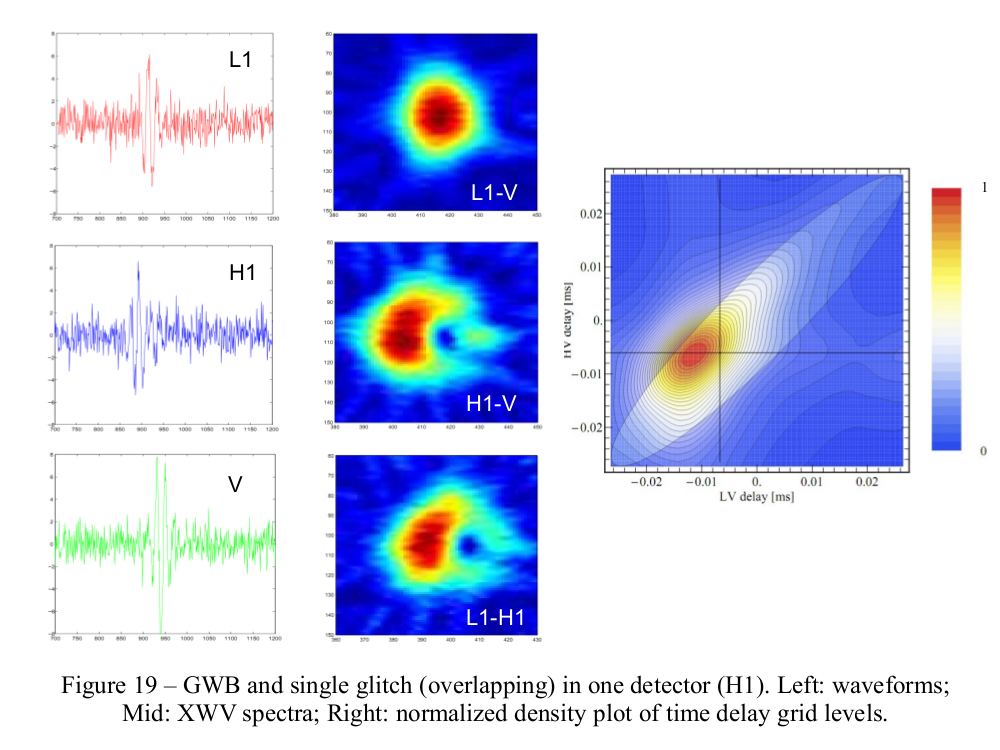}}
\end{figure}
\begin{figure}
\centerline{\includegraphics[scale=0.5]{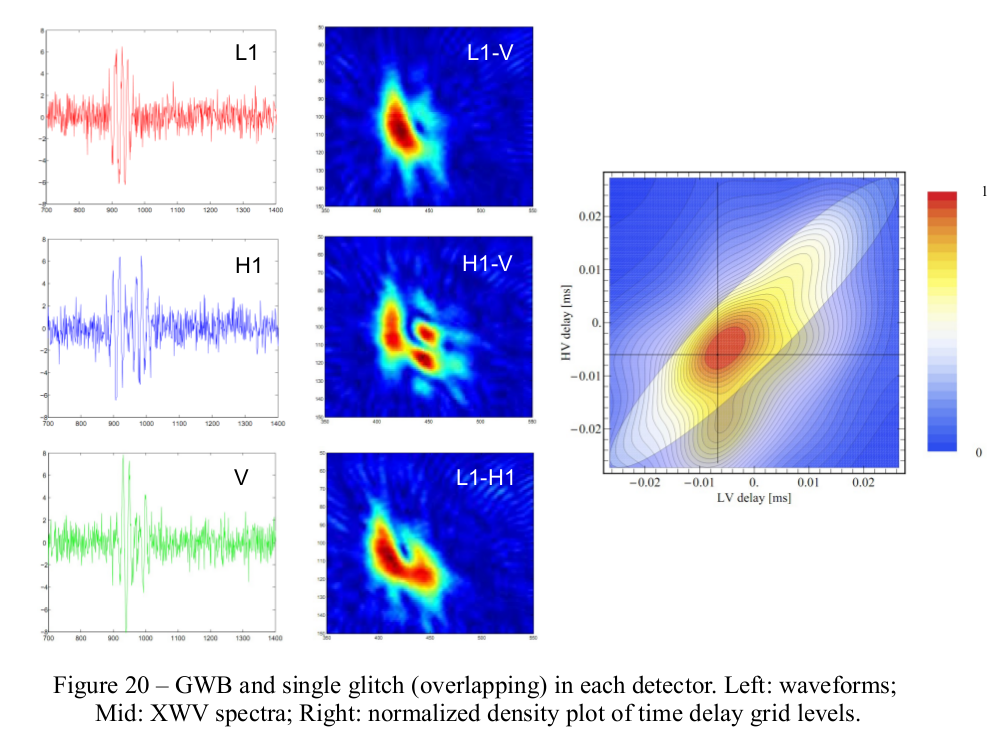}}
\end{figure}

\end{document}